\documentclass[entropy,article,accept,pdftex,moreauthors]{Definitions/mdpi} 

\firstpage{1}
\pubvolume{1}
\issuenum{1}
\articlenumber{0}
\pubyear{2022}
\copyrightyear{2022}
\datereceived{} 
\dateaccepted{} 
\datepublished{} 
\hreflink{https://doi.org/} 

\usepackage{amsmath,amsfonts,amssymb,amsthm}
\usepackage{graphicx}
\usepackage{color}
\usepackage{subfigure}

\newcommand{\ket}[1]{ | \, #1 \rangle} \newcommand{\bra}[1]{ \langle #1 \, |} 
\newcommand{\proj}[1]{\ket{#1}\bra{#1}}

\newcommand{\TN}[1]{ \left| \left| #1 \, \right| \right|_\text{Tr}} 
\newcommand{\Ab}[1]{ \left| #1 \, \right|} 
\newcommand{\euler}{e}

\newcommand{\iu}{{i\mkern1mu}}

\DeclareMathOperator{\Tr}{Tr}

\DeclareRobustCommand\openone{\leavevmode\hbox{\small1\normalsize\kern-.33em1}}%

\Title{Non-perfect propagation of information to noisy environment with self-evolution}

\TitleCitation{Non-perfect propagation of information}

\Author{Piotr Mironowicz $^{1,2}$*\orcidA{}, Paweł Horodecki $^{1,3}$\orcidB{} and Ryszard Horodecki $^{1}$\orcidC{}}

\address{%
	$^{1}$ \quad International Centre for Theory of Quantum Technologies, University of Gda\'{n}sk, Wita Stwosza 63, 80-308 Gda\'{n}sk, Poland\\
	$^{2}$ \quad Department of Algorithms and System Modeling, Faculty of Electronics, Telecommunications and Informatics, Gda\'{n}sk University of Technology, Gabriela Narutowicza 11/12, 80-233 Gda\'{n}sk, Poland\\
	$^{3}$ \quad Faculty of Applied Physics and Mathematics, Gda\'{n}sk University of Technology, Gabriela Narutowicza 11/12, 80-233 Gda\'{n}sk, Poland
}
\corres{Correspondence: piotr.mironowicz@gmail.com}

\abstract{We study the non-perfect propagation of information to evolving low-dimensional environment that includes self-evolution as well as noisy initial states and analyze interrelations between the degree of objectivization and environment parameters. In particular, we consider an analytical model of three interacting qubits and derive its objectivity parameters. The numerical analysis shows that the quality of the  spectrum broadcast structure formed during the interaction may exhibit non-monotonicity both in the speed of self-dynamics of the environment as well as its mixedness. The former effect is particularly strong, showing that - considering part of the environment as a measurement apparatus - an increase of the external magnetic field acting on the environment may turn the very vague measurement into close to ideal. 
	The above effects suggest that quantum objectivity may appear after increasing the dynamics of the environment, although not with respect to the pointer basis, but some other one which we call generalized pointer or indicator basis. 
	Furthermore, it seems also that when the objectivity is poor it may be improved, at least by some amount, by increasing thermal noise. We provide further evidence of that by analyzing the upper bounds on distance to the set of states representing perfect objectivity in the case of a higher number of qubits. 
	}


\begin{document}
	
	\section{Introduction}
	
	Quantum mechanics works perfectly and is reliable in an appropriate regime. Nevertheless, it leaves us with cognitive discomfort as the theory which pretends to be fundamental should describe whole physical reality including classical objective properties of the systems that are inter-subjectively verifiable by independent observers. The problem is that quantum formalism does not offer a simple footbridge from the quantum world to our actual world. This issue involves many aspects, it has a long history and huge literature~\cite{Lands}. In particular, it involves a highly non-trivial question: Is it possible to get around the fundamental restrictions  (no-broadcasting~\cite{NoncomBarnum, noBrodcastPiani}) on the processing of quantum information to explain the emergence of the objective nature of information redundancy in the actual world?
	
	Thanks to Zurek's quantum Darwinism concept~\cite{Zurek 2009} there are strong reasons to believe that the decoherence theory pioneered by Zeh~\cite{Zeh70} and developed by Zurek~\cite{Zurek81, ZurekRMP} and others~\cite{Joos2003, SchlosshauerRMP} based on the system-environment (or, in the Bohr's spirit: system- context~\cite{Grangier}) paradigm offers the most promising approach to the emergence of classicality from the quantum world.
	
	Quantum Darwinism (QD) considers a decohering environment $E$ as a ``witness'' that monitors and can reveal the information about a system $\mathcal{S}$. The environment consists of multiple independent N fragments and objectivity emerges when interacting with the system led to redundant information proliferation about system $\mathcal{S}$ measured by quantum mutual information $I(\mathcal{S}:\mathcal{E})$ between system and an accessible fragment of the environment, $\mathcal{E} \subset E$, where $I(\mathcal{S}:\mathcal{E}) = H(\mathcal{S}) + H(\mathcal{E}) - H(\mathcal{S}\mathcal{E})$ is the mutual information between the system and part of the environment and $H(\cdot)$ is the von Neumann entropy (see~\cite{Zurek_eavesdropping} and ref. therein). The term ``objectivity'' means that the state of the system satisfies the following:
	\begin{Definition}
		\label{dfn:Objective}
		A system state $\mathcal{S}$ is objective when many independent observers can determine the state of S independently, without perturbing it, and arrive at the same result~\cite{ EnvirWitness, Qorigins, MonitMironowicz}.
	\end{Definition}
	
	The different theoretical and experimental implementations of QD has been considered and discussed based on information-theoretic condition:
	\begin{equation}
		\label{eq:ITC}
		I(\mathcal{S}:\mathcal{E}) = H(\mathcal{S}).
	\end{equation}
	 In many cases, the above relation is enough to identify correctly emergent objective properties in a quantum system in contact with an environment. Interestingly, sometimes the nature of the quantum-classical interplay may be richer. In particular, examples have been found in which QD can falsely announce objectivity and it has been pointed out that QD can be inconsistent with the emergence of objectivity when the condition~\eqref{eq:ITC} is used ~\cite{Qorigins,QD_struct.envir,Le_QD_SBS}.
	
	In connection with the QD, a problem arises: \textit{To identify quantum primitive information broadcasting state responsible for the emergence of the perceived objectivity}. This issue was raised in~\cite{Qorigins} where it was proved  that Bohr's non-disturbance measurement, full decoherence, and ``strong independence'' lead to the paradigmatic spectrum broadcast structure (SBS) responsible for objectivity, which can be written in the following form:
	\begin{equation}
		\varrho_{\mathcal{S} \mathcal{E}} = \sum_i p_i \proj{\psi_i} \otimes \varrho_i^{\mathcal{E}_1} \otimes \cdots \otimes \varrho_i^{\mathcal{E}_N},
		\label{SBS-canonical}
	\end{equation}
    where $\mathcal{E}$ is the accessible environment, $\mathcal{E}_k \in \left\{ \mathcal{E}_1, \mathcal{E}_2, \dots, \mathcal{E}_N \right\}$, $\mathcal{E}_{k} \cap \mathcal{E}_{k'} = \emptyset$, $\mathcal{E}_k \subset \mathcal{E}$ are the subenvironments. The conditional states $\{ \varrho_i^{\mathcal{E}_k} \}$ can be used to perfectly distinguish index $i$, where $\{\ket{\psi_i}\}$ is some diagonal basis of the $\mathcal{S}$ and $\{p_i\}$ its spectrum. 
	
    The basis $\{\ket{\psi_i}\}$ has a special role in the above picture. It represents the objective information about the quantum system. The above form (\ref{SBS-canonical}) is agnostic about the physical mechanism leading to it. Hence, we shall call the basis $\{\ket{\psi_i}\}$ an {\it generalised pointer basis} (or, alternatively {\it indicator basis}). In the case of quantum Darwinism, when determined by the interaction Hamiltonian, this basis becomes exactly the pointer basis. However, there may be other physical processes that lead to the above  (\ref{SBS-canonical}) structure. This is directly related to the main point of the present paper: any pointer basis is the generalized pointer basis, but {\it not} vice versa.
	
	The above SBS state clearly shows the meaning of the terms ``objective''/ inter-subjective used in Definition~\ref{dfn:Objective}. It reveals the contextual nature of objectivity which emerges as a property of a system dependent on the combined properties of the system and the environment. These states have a discord zero hence only the ``classic'' spectrum of the system $\{p_i\}$ is broadcast to the environment and therefore independent observers do not have access to quantum information. It has been proved that SBS is a stronger condition than the QD, i.e. SBS implies QD ~\cite{Qorigins}. The objective states with spectrum broadcast structure can be used as ideal ``frames of reference'' to which any real states can be compared. The SBS was identified in the many models of open quantum systems (see~\cite{Korbicz_road} and ref. therein) and its simulations on a quantum computer were demonstrated~\cite{Palma_WitnessObjComputer}. It was also shown that the objectivity is subjective across quantum reference frames~\cite{Blurred} including its dynamical aspects~\cite{Tuziemski}.

    It was mentioned in~\cite{Qorigins} that the SBS-like states may open a ``classical window'' for life processes within the quantum world.  Interestingly the process of objectivization of information over time was analyzed using quantum state discrimination and potential applications for the theory of evolution of senses were suggested~\cite{MonitMironowicz}.  Remarkably, in nature, there are thermal states the properties of which, seem to contradict objectivity suggesting that thermality and objectivity are mutually exclusive. Recently Le et al.~\cite{Winter_Thermality} examined the overlap between thermal and objective states and showed, that there are certain regimes in which exist states that are approximately thermal and objective.

    As was mentioned above, the SBS implies quantum Darwinism condition~\eqref{eq:ITC}, however, the opposite implication does not hold. The discrepancy between the QD and SBS led to the discovery of a stronger version of quantum Darwinism (SQD)~\cite{SQD}, where~\eqref{eq:ITC} is replaced by a stronger condition: A system state  is objective iff the following conditions hold simultaneously:
	\begin{subequations}
		\begin{equation}
			I(\mathcal{S}:\mathcal{E}) = \chi(\mathcal{S}:\mathcal{E}),
		\end{equation}
		\begin{equation}
			I_\text{acc}(\mathcal{S}:\mathcal{E}_k) = H(\mathcal{S}),
		\end{equation}
		\begin{equation}
			I(\mathcal{E}_1 \cdots \mathcal{E}_N | \mathcal{S}) = 0,
		\end{equation}
	\end{subequations}
	where 
	$\chi(\mathcal{S}:\mathcal{E})$ is the Holevo information in the pointer basis $\pi$, $I_\text{acc}(\mathcal{S}:\mathcal{E}_k)$ is the accessible information, and $I(\mathcal{E}_1 \cdots \mathcal{E}_N | \mathcal{S})$ is the conditional multipartite mutual information. It has been shown that SQD is equivalent to bipartite SBS and it is sufficient and necessary for objectivity~\cite{SQD,CommentSQD,ReplaySQD}. Thus SBS and SQD are the two extensions  of the standard QD  based on the quantum state structure and information respectively~\cite{Witnessing-non-object}.
	
	However, in the limit of a large environment the standard QD works very well. Namely, in \cite{Zurek_eavesdropping}  the authors investigated a model based on imperfect C-NOT gates and showed that relevant quantities for QD exhibit similar dependence on the size $\Ab{\mathcal{E}_k}$ of a fragment of environment $\mathcal{E}_k$ including scaling independent from the quality of the imperfect C-NOT gates and the size of the fragment of environment $\mathcal{E}_k$.
    
	\section{Aspects of emergence of objective information on quantum ground}
    
    The fundamental elements of Zurek’s quantum Darwinism discovery were 
    (1) the methodological identification that classical correlations between the system and environment and redundant character of the information about the system in the environment are a constitutive feature of objectivity 
    (2) proof that this objective information is very special, unambiguously determined by a system-environment interaction. More precisely, the interaction chooses a basis, called the pointer basis, and this is the information concerning the question ``In which state of the pointer basis the system is?'' that is replicated by interaction in the environment in a stable way. Quite remarkably, the latter feature is responsible for the strong cognitive power of the whole process. 
    
    This is the case for the following three reasons. First, the information-theoretic correlations between the system and parts of the environment have a classical, well-understood character. Second, a subject observing a part of the environment not only knows that the system is in some particular state but also knows exactly what the system state is, since the latter belongs to a special basis - the pointer basis. Third, by a repetition of an experiment of putting the system in the same state many times into the environment and observing some part of the latter, the subject is also able to learn (via a collection of the experiment statistics) about some parameters of the initial state of the system. They correspond to the diagonal of the state written in the pointer basis. Those parameters are just revealed in this process. In this sense, we may understand the quantum Darwinism process as a process of objectivization that discloses parameters of the system state. 
        
    In the present paper, we inquire as to whether and when the dynamical emergence of objectivity is possible in a more relaxed sense, namely, when one retains only the element (1) of Zurek’s program. More precisely, we only demand that the information about the system being in one of the elements of some basis is classically present in the environment - there are only classical correlations between system and environment. However neither the basis need to be directly related to the system-environment interaction nor the corresponding statistics need to directly correspond to some particular parameters of the initial state of the system. In this sense, the basis has only the character of the generalized pointer basis  (see discussion below \eqref{SBS-canonical}).
        
       
    Below we will show that this kind of objectivization can emerge in low-dimensional qubit systems. For this purpose, we examine the non-perfect propagation of information from system $\mathcal{S}$ to the noisy environment $\mathcal{E}$ with self-evolution and analyze interrelations between the degree of objectivization and environment parameters. We consider two different environments, the first composed of one observed and one unobserved qubit and the second one where there are seven observed qubits and one unobserved.
    
    In particular, we consider an analytical model of three interacting qubits and derive its objectivity parameters. Then we show that if the imperfection of the C-NOT gate is known, the emergence of the objectivity albeit with respect to a different basis than the one associated with the gate itself - can be triggered by a carefully chosen environment self-dynamics. For a seven-qubit environment, numerical calculations show that dynamics of the environment may help the emergence of relaxed objectivity to happen.

	\section{Analytical model for three interacting qubits}
	\label{sec:analytical}
	
	Let us now investigate a model of three interacting qubits, where we consider one of them as the observed system, and the remaining two constitute the observing environment $\mathcal{E}$. In the following we will derive a closed analytical formula for the objectivity parameters, \textit{viz. }decoherence and orthogonalization, in a scenario where the information is widespread using imperfect C-NOT gate (C-INOT gate), and where the time evolution includes self-evolution of each of the qubits and their inter-environmental interaction.
	
	\subsection{Derivation of objectivity parameters}
	
	We model the  C-INOT gates~\cite{Zurek_eavesdropping} 
	defined by the formula:
	\begin{equation}
		\label{eq:Uc-maybe}
		U_\text{C-INOT} \equiv
		\begin{bmatrix}
			1 & 0 & 0 & 0 \\
			0 & 1 & 0 & 0 \\
			0 & 0 & \sin(\theta) & \cos(\theta) \\
			0 & 0 & \cos(\theta) & -\sin(\theta)
		\end{bmatrix},
	\end{equation}
	where $\theta \in [0,\pi / 2]$ is the imperfection parameter. Note that for $\theta = 0$ the gate reproduces the perfect C-NOT gate. It doesn't allow to model the two qubit identity unitary. In this work we have chosen the Kronecker product convention where the primal structure of the matrix representation is determined by the first space involved in the product.
	
	There is an infinite number of Hamiltonians that can realise the gate~\eqref{eq:Uc-maybe} after some fixed time of interaction. Here we choose the following Hamiltonian:
	\begin{equation}
		\label{eq:Hc-maybe}
		H_\text{C-INOT} \equiv
		\begin{bmatrix}
			0 & 0 & 0 & 0 \\
			0 & 0 & 0 & 0 \\
			0 & 0 & (\pi / 2) (1-\sin(\theta)) &  -(\pi / 2) \cos(\theta) \\
			0 & 0 & -(\pi / 2) \cos(\theta) & (\pi / 2) (1+\sin(\theta))
		\end{bmatrix}.
	\end{equation}
	One may check that $\exp(-i t H_\text{C-INOT}) = U_\text{C-INOT}$ for $t=1$. We denote by $H_\text{C-INOT}^1$, $H_\text{C-INOT}^2$, the Hamiltonians of C-INOT acting on the first and second qubits of the environment, respectively, conditioned by the system bit.
	
    We assume that the total Hamiltonian is given by:
	\begin{equation}
		\label{eq:H_total}
		H_\text{TOTAL} \equiv H_\text{C-INOT}^1 + H_\text{C-INOT}^2 + \alpha_1 H_1 + \alpha_2 H_2 + \alpha_3 H_3.
	\end{equation}
	where 
	\begin{subequations}
		\label{eq:Hs}
		\begin{equation}
			H_1 \equiv \sigma_Z \otimes \openone_2 \otimes \openone_2,
		\end{equation}
		\begin{equation}
			\label{eq:H_2}
			H_2 \equiv \openone_2 \otimes \sigma_Z \otimes \openone_2 + \openone_2 \otimes \openone_2 \otimes \sigma_Z,
		\end{equation}
		\begin{equation}
			\label{eq:H_3}
			\begin{aligned}
				H_3 \equiv & \sigma_Z \otimes \sigma_Z \otimes \openone_2 + \sigma_Z \otimes \openone_2 \otimes \sigma_Z + \\
				& \openone_2 \otimes \sigma_Z \otimes \sigma_Z + \sigma_Z \otimes \sigma_Z \otimes \sigma_Z,
			\end{aligned}
		\end{equation}
	\end{subequations}
	with $\openone_2$ denoting identity on a single qubit space. Here $H_1$ is the self-evolution Hamiltonian of the central system; $H_2$ is the self-evolution of the environmental qubits, that can be e.g. caused by an external magnetic field; $H_3$ contains inter-qubit interactions, between each pair of the qubits plus the joint interaction between all three qubits via $ZZ$ and $ZZZ$ coupling, respectively. $\alpha_1,\alpha_2,\alpha_3 \geq 0$ are the interaction strength parameters.
	
    Note that whereas~\eqref{eq:H_2} easily generalizes for cases with more qubits in the environment, but~\eqref{eq:H_3} is specific for the two-qubit case. Further in this paper, we consider other inter-environmental interactions with larger environments.
	
    One can rewrite $H_\text{TOTAL} = (\pi - \alpha_1) \openone_8 + M$ with $\openone_8$ being the 3-qubit identity operator, and $M$ a block-diagonal matrix, giving $V \equiv \exp(-i t M)$ also of block-diagonal form, with blocks denoted by $V_0$ and $V_1$. The explicit form of those matrices is given in Appendix~\ref{sec:AppA}.
	
	We assume that the initial system-environment state $\varrho_{S\mathcal{E}}$ is given by:
	\begin{equation}
		\varrho_{S\mathcal{E}} = \proj{+} \otimes \varrho^{\mathcal{E}_1} \otimes \varrho^{\mathcal{E}_2},
	\end{equation}
	where
	\begin{equation}
		\label{eq:initMixp}
		\varrho^{\mathcal{E}_1} = \varrho^{\mathcal{E}_2} = \varrho \equiv p \proj{0} + (1-p) \proj{1}
	\end{equation}
	are the environment qubit states, $p \in [0,0.5]$.
	
	We note that for $\alpha_1 = \alpha_3 = 0$ the state $\varrho$ is the termal state of the environment and $p = \frac{\euler^{-\alpha_2/\beta}}{\euler^{-\alpha_2/\beta}+\euler^{\alpha_2/\beta}}$, or $1/\beta = \frac{1}{2 \alpha_2} \ln ((1-p)/p)$, where $\beta$ is the inverse temperature. This holds because of the form of $H_2$, i.e. the state $\ket{1}$ is the ground state of the Hamiltonian. After the time evolution, given by $\exp(-i t H_\text{TOTAL})$, the joint state of the system and two qubit environment {\it in the computational basis} of the observed system is
	\begin{equation}
		\rho_{S \mathcal{E}_1 \mathcal{E}_2 \text{comp}} =	(1/2) 
		\begin{bmatrix}
			V_0 E V_0^\dagger & V_0 E V_1^\dagger \\
			V_1 E V_0^\dagger & V_1 E V_1^\dagger
		\end{bmatrix},
	\end{equation}
	where each element of the 2x2 matrix is a block 4x4 matrix and $E \equiv \varrho \otimes \varrho$ and $V_k \equiv \exp(-i t M_k)$, k=0, 1. After tracing out the second environmental qubit we get, again, in the computational basis, the following two qubit joint state of the system and observing qubit:
	\begin{equation}
		\label{eq:compState}
		\rho_{S \mathcal{E}_1 \text{comp}} = (1/2) 
		\begin{bmatrix}
			\Tr_2(V_0 E V_0^\dagger) & \Tr_2(V_0 E V_1^\dagger) \\
			\Tr_2(V_1 E V_0^\dagger) & \Tr_2(V_1 E V_1^\dagger)
		\end{bmatrix},
	\end{equation}
	where $\Tr_2$ is the second qubit partial trace operation. Hence, we obtain the collective decoherence factor in the form
	\begin{equation}
		\label{eq:Gamma}
		\Gamma = \TN{\Tr_2 \left( V_0 E V_1^\dagger \right) }.
	\end{equation}
	This equation is the value of the trace norm of a 2x2 upper off-diagonal block of the 4x4 matrix \eqref{eq:compState}. The trace norm is defined as $\TN{A} = Tr( \sqrt{A^{\dagger} A})$.

	The probabilities $c_0$ and $c_1$ of the system to be in a state $0$ or $1$ of the computational basis are given by $c_i = (1/2) \Tr{(V_i E V_i^\dagger)} = 0.5$ and reveal to be constant in time. Conditioning upon the system state in the computational basis and tracing out the second environmental qubit, we get that the conditional states of the remaining (observing) qubit, denoted $\varrho_0$ and $\varrho_1$, where $\varrho_i \equiv \bra{i}_S \left( \rho_{S \mathcal{E}_1} \right) \ket{i}_S$ is a single qubit. Those states are obtained by a projection of the joint state of the system and part of the environment on one of the possible states of the system in the computational basis. If there is no coherence between different states of the system, then the off-diagonal elements should vanish, as is explicitly stated in the definition of SBSs.
	
    The (generalized) fidelity~\cite{FuchsFidelity} (also called the Bhattacharyya coefficient), used as a measure of state overlap~\cite{MonitMironowicz} for two matrices $\varrho_0$ and $\varrho_1$ is defined as
	\begin{equation}
		\label{eq:B}
		\mathcal{F}(\varrho_0, \varrho_1) \equiv \Tr \sqrt{\sqrt{\varrho_0} \varrho_1 \sqrt{\varrho_0}}.
	\end{equation}
	The larger the value of the fidelity, the poorer is the orthogonalization of the relevant observable.
	We provide explicit formulae for~\eqref{eq:Gamma} and~\eqref{eq:B} in Appendix~\ref{sec:AppB}.

	The upper bound to the distance to the Spectrum Broadcast Structure~\cite{Qorigins,MonitMironowicz} is
	\begin{equation}
		\label{eq:OurPRLUpperBound}
		||\rho_{S \mathcal{E}_1} - \rho^{(SBS)}_{S \mathcal{E}_1}|| \leq 2 \left( \Gamma + \sqrt{c_0 c_1} \mathcal{F}(\varrho_0, \varrho_1) \right)
	\end{equation}
	which is true for some state $\rho^{(SBS)}_{S \mathcal{E}_1}$ having the SBS form~\eqref{SBS-canonical}. The bound~\eqref{eq:OurPRLUpperBound} can be applied to any state, not only qubit-qubit states.  In Appendix~\ref{sec:AppC} we discuss the distance of evolved state to thermal state.
	
	\subsection{ Generalised pointer basis optimal for SBS}
	\label{sec:optBasis}
	
    Since the constituent Hamiltonians in~\eqref{eq:Hs} do not commute with the C-INOT gate Hamiltonians, one cannot follow the paradigm of~\cite{Zurek81} and determine the generalized pointer basis from the interaction Hamiltonian only. In other words, this is the case when the generalized pointer basis (that may be also called indicator basis) is a different object than the pointer basis known from quantum Darwinism.
	
    Above in~\eqref{eq:compState} we wrote the evolved state in the computational basis, and the calculations of~\eqref{eq:obss} and~\eqref{eq:Bclosed} refer to this basis. On the other hand, one may ask the question, whether there exists some other basis of the observed system, that manifests structure closer to SBS.
	
    For the two environmental qubit cases with one of them being traced out we shall look for the optimal  SBS state, namely, the one that is the closest to the actual system-environment state represented in the computational basis $\rho_{S \mathcal{E}_1 \text{comp}}$ (see \eqref{eq:compState}).  To this aim we shall minimize the distance of the latter to the SBS states which, by the very definition, have the form:
	\begin{equation}
		\label{eq:optState}
	 \rho_{S \mathcal{E}_1 }^{SBS} = \tilde{p} \proj{\psi}^S \otimes \proj{\chi}^{\mathcal{E}_1} + (1-\tilde{p}) \proj{\psi^\perp}^S \otimes \proj{\chi^\perp}^{\mathcal{E}_1}.
	\end{equation}
	Note that this form easily generalizes for environments of higher dimension with $\proj{\chi}^{\mathcal{E}_1}$ and $\proj{\chi^\perp}^{\mathcal{E}_1}$ replaced with orthogonal $\varrho_0^{\mathcal{E}_1}$ and $\varrho_1^{\mathcal{E}_1}$:
	\begin{equation}
		\label{eq:optStateHigher}
		\rho_{S \mathcal{E}_1 }^{SBS} = \tilde{p} \proj{\psi}^S \otimes \varrho_0^{\mathcal{E}_1} + (1-\tilde{p}) \proj{\psi^\perp}^S \otimes \varrho_1^{\mathcal{E}_1}.
	\end{equation}
    In the case of two-qubit case minimisation of the corresponding distance 
    \begin{equation}
		\label{eq:targetOptBasis}
		\TN{\rho_{S \mathcal{E}_1 \text{comp}} - \rho_{S \mathcal{E}_1}^{SBS}}
	\end{equation}
	over all probability $\tilde{p}$, and state vectors $\ket{\psi}$, and $\ket{\chi}$ defining (\ref{eq:optState}) gives the optimal  SBS state:
	\begin{equation}
		\label{eq:optState-1}
	 \rho_{S \mathcal{E}_1 \text{opt}}^{SBS} = \tilde{p}^{*} \proj{\psi^{*}}^S \otimes \proj{\chi^{*}}^{\mathcal{E}_1} + (1-\tilde{p}^{*}) \proj{\psi^{* \perp}}^S \otimes \proj{\chi^{* \perp}}^{\mathcal{E}_1}.
	\end{equation}

	The basis $\{\ket{\psi^{*}}, \ket{\psi^{*\perp}}\}$ for which the minimum of~\eqref{eq:targetOptBasis} is attained should be considered as a candidate for the generalised pointer  (equiv. indicator) basis for the case when the total Hamiltonian~\eqref{eq:H_total} does not commute with the interaction Hamiltonians~\eqref{eq:Hc-maybe}.
	
    To be more specific, for $\ket{\psi}$ and $\ket{\chi}$ being qubits, as in~\eqref{eq:optState}, we use the standard Bloch parametrization
	\begin{equation}
	    \label{eq:BlochParametersSBS}
		\ket{\psi} = \cos \left( x_\psi / 2 \right) \ket{0} + \sin \left( x_\psi / 2 \right) \exp \left( i y_\psi \right)\ket{1},
	\end{equation}
	with $x_\psi \in [0, \pi]$, $y_\psi \in [0, 2 \pi]$, and similarly for $\ket{\chi}$. Further, without loss of generality, we assume $\tilde{p} \in [0.5, 1]$. This last assumption assures continuity of the parameters obtained in the optimization, as without it the optimization has two possible equivalent solutions, \textit{viz.} the one from~\eqref{eq:optState} and the second one with $\tilde{p}$ replaced with $1-\tilde{p}$ and states replaced with their orthogonal complements.
	
\begin{figure}
		\centering   
		\subfigure[SBS distance]{\label{fig:illustr_dist}\includegraphics[width=60mm]{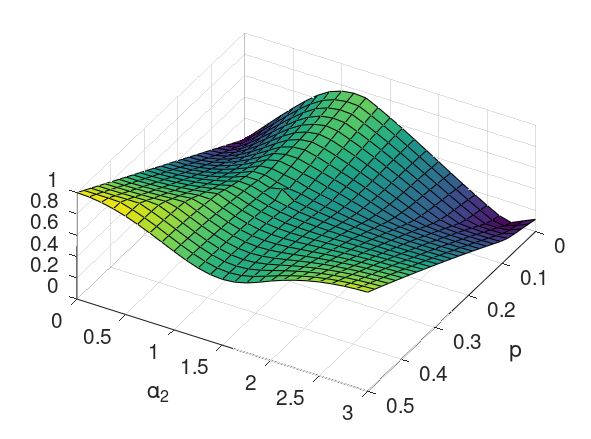}}
		\subfigure[$\tilde{p}$]{\label{fig:illustr_p}\includegraphics[width=60mm]{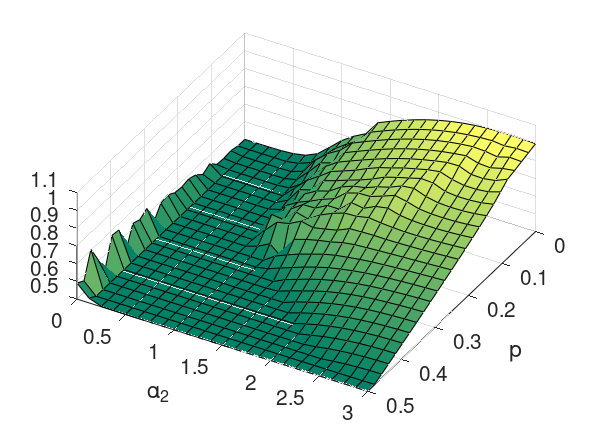}}
		
		\subfigure[$\max(\cos(x_\psi/2), \sin(x_\psi/2))$]{\label{fig:illustr_theta}\includegraphics[width=60mm]{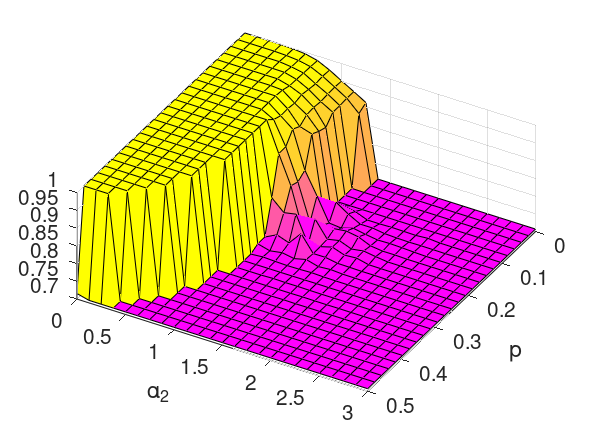}}
		\subfigure[$y_\psi$]{\label{fig:illustr_phi}\includegraphics[width=60mm]{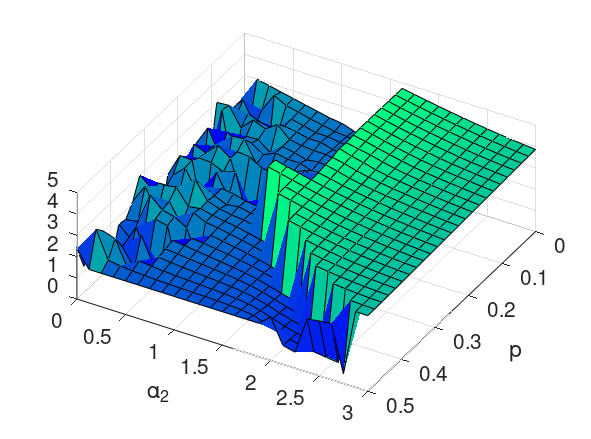}}
		\caption{\label{fig:illustr} Sample results of SBS basis optimization~\eqref{eq:optState} using the Bloch parametrization~\eqref{eq:BlochParametersSBS}. We consider the state state after time $t = 1$, with Hamiltonian~\eqref{eq:H_total} with parameters $\alpha_1 = \alpha_3 = 0$, for different values of $\alpha_2$ and environmental mixedness $p$, cf.~\eqref{eq:initMixp}, and perfect CNOT interaction. Figure~\ref{fig:illustr_dist} contains the minimized distance~\eqref{eq:targetOptBasis} obtained for $\tilde{p}$, $x_\psi$ and $y_\psi$ parameters shown at Figs~\ref{fig:illustr_p}, \ref{fig:illustr_theta}, and~\ref{fig:illustr_phi}, respectively. Note that Fig.~\ref{fig:illustr_dist} is the same as in Fig.~\ref{fig:2qa} (seen from a different angle). For $x_\psi$ in Fig.~\ref{fig:illustr_theta} we used trigonometric transformation, so that the value $1$ refers to the computational basis. Note that the phases factor $y_\psi$ of the Bloch qubit strongly fluctuates in the region where the computational basis is optimal, as in that case $y_\psi$ has no impact on the state.  The Fig. \ref{fig:illustr_theta} the yellow part corresponding to standard basis and the light purple one representing  bases complementary to the standard basis. The latter bases are in general different from Hadamard basis, which can be seen by examination of the phases in Fig.\ref{fig:illustr_phi}. Each of the basis in the light purple region represents some {\it generalised pointer basis} (see the discussion at the beginning of section 3.2). 
		}
	\end{figure}

    In the actual numerical calculations, we used unconstrained gradient search with a continuous map $\mathbb{R} \to [0.5,1]$ for the $\tilde{p}$ parameter, and postprocessing of the resulting optimal values $x_\psi, y_\psi, x_\chi, y_\chi \in \mathbb{R}$ to obtain angles withing the proper Bloch parameter range yielding the same qubit states. We illustrate the optimization of the SBS basis at Fig.~\ref{fig:illustr}.
	
	\subsection{Marginal cases}
	\label{sec:marginal}
	
	Another interesting marginal case is for maximally mixed environment, \textit{i.e.} for $p = 0.5$. Then $\mu = 1/4$ and $\nu = 0$, again leading to $\mathcal{F}(\varrho_0, \varrho_1) = 1$. This is in agreement with~\cite{Zurek09}, as this case refers to maximal entropy of the environment, and thus its capacity is $0$.
	
	For fixed $p \neq 0.5$ and $\theta < \pi / 2$ we see from~\eqref{eq:Bclosed} that the orthogonalization factor is a function of $r_4$. Thus, by changing the difference $\alpha_2 - \alpha_3$ we can adjust the total Hamiltonian so that the orthogonalization reaches its maximum.
	Thus, knowing imperfections of the interaction $\theta$, mixedness $p$ environment, and the internal interaction $H_3$, we can \textit{e.g.} manipulate the magnetic field $H_2$ acting on the environment, to improve the quality of the measurement. We illustrate this adjustment in the following section.

	\section{Central Interaction: Optimization of Spectrum Broadcast Structure for $2$ environmental qubits}
	
	We now consider the case when $\alpha_1 = \alpha_3 = 0$, and $\alpha_2, \theta \geq 0$, \textit{i.e.} with imperfect central interaction and self-evolution of environmental qubits with initial mixedness parameter $p$ after time $t = 1$, \textit{viz.} at the time after which the central interaction has fully occured.
	
    We first note that in the former section~\ref{sec:analytical} we considered the Spectrum Broadcast Structure obtained in the pointer basis~\cite{Zurek81}, which was in that case equal to the computational basis of the observed system. Yet, it is possible to calculate the SBS distance for different basis, \textit{viz.} for the optimal basis, as introduced in sec.~\ref{sec:optBasis}. We used the gradient method~\cite{gradientMethod} to find the basis that minimizes the SBS distance. We note that the considered setup with only two qubits is very far from the one involving the macroscopic environment and thus the objectivity present in this model can be only temporary since a single qubit is not able to induce full decoherence that is stable in time or orthogonalization of observables.
	
	
    Yet, here we are interested in the classical properties of the evolved system at a particular time moment, namely the time that we denote as $t=1$, the time at which the measurement is supposed to occur. Still, this scenario illustrates the mechanism, and we leave the actual scaling of the discussed non-monotonic phenomena for further research.
	
	\begin{figure}
		\centering   
		\subfigure[$\theta = 0$]{\label{fig:2qa}\includegraphics[width=60mm]{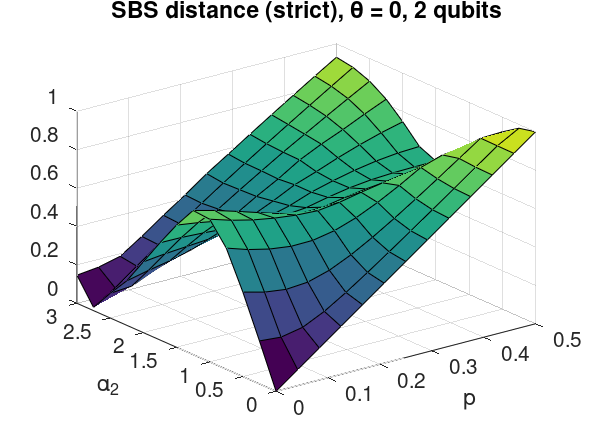}}
		\subfigure[$\theta = \pi / 8$]{\label{fig:2qb}\includegraphics[width=60mm]{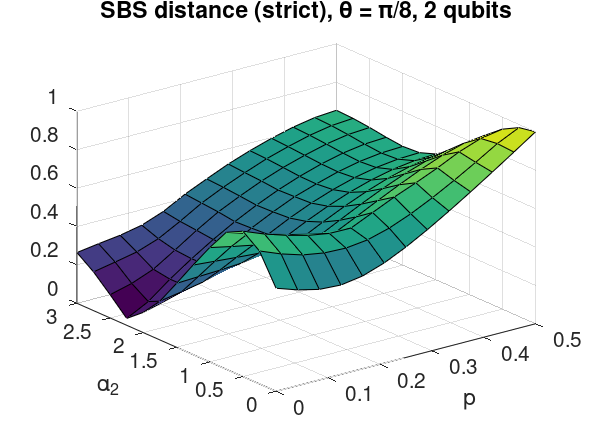}}
		
		\subfigure[$\theta = \pi / 4$]{\label{fig:2qc}\includegraphics[width=60mm]{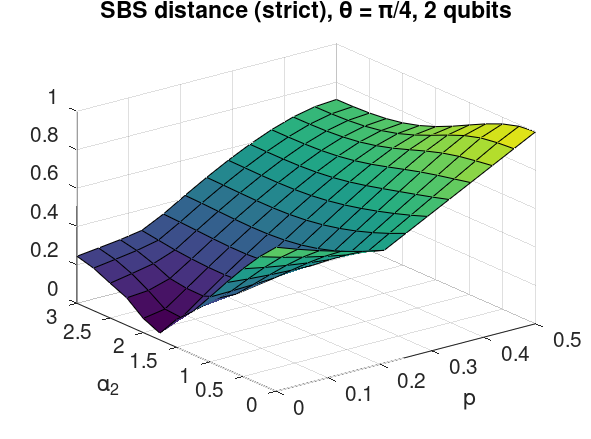}}
		\subfigure[$\theta = 0.9 \pi / 2$]{\label{fig:2qd}\includegraphics[width=60mm]{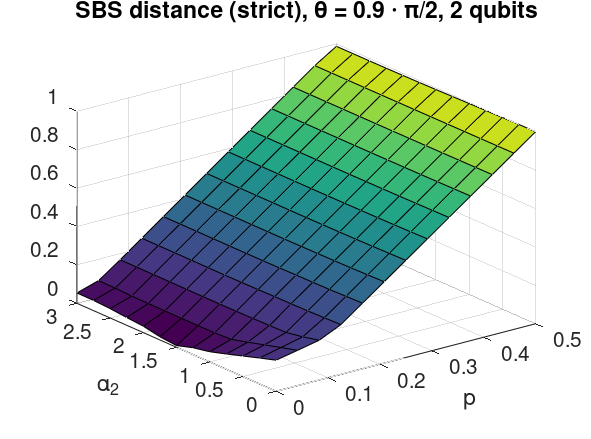}}
		\caption{\label{fig:2qs}SBS distance for C-INOT central interaction with various values of the gate imperfection parameter $\theta$ with $2$ environmental qubits. Each value of $\theta = 0, \pi/8, \pi/4, 0.9 \pi / 2$ refers to different interaction between the central system and each of the environmental qubits, as given in~\eqref{eq:Uc-maybe}. The axis $\alpha_2$ describes the strength of the self-evolution of the environmental qubits, see~\eqref{eq:H_2}, and $p$ refers to the initial mixedness of the environmental qubits, see~\eqref{eq:initMixp}. The figure illustrates non-monotonic dependence of the distance of the evolved state from the closes SBS state of the form~\eqref{eq:optState} from the parameters $\alpha_2$ and $p$. In particular, it can be seen that in many cases it is not the smallest value of mixedness, that leads to states close to the SBS form, but the ``optimal'' environment mixedness $p$ depends on the value of the self-evolution strength $\alpha_2$.}
	\end{figure}
	
\begin{figure}
		\centering
		\subfigure[Dependence of the SBS distance as a function of $\alpha_2$ for $\alpha_1=\alpha_3=p=0$ for various values of $\theta$.]{\label{fig:2q_alpha2s}\includegraphics[width=55mm]{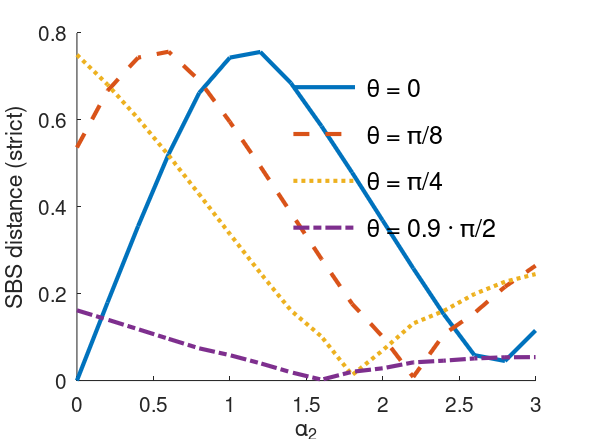}}
		\qquad
		\subfigure[Dependence of the SBS distance as a function of $p$ for $\alpha_1=\alpha_2=\alpha_3=0$ for various values of $\theta$.]{\label{fig:2q_ps}\includegraphics[width=55mm]{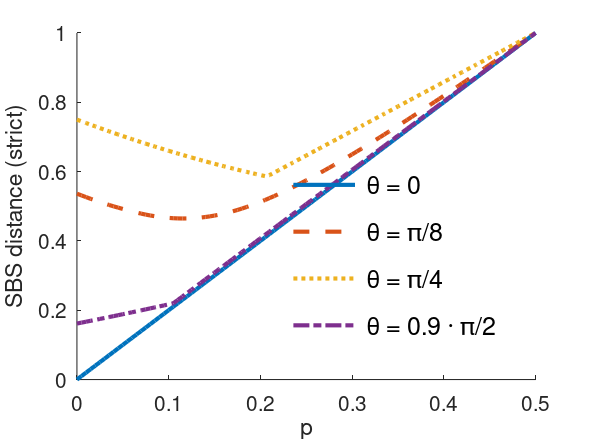}}
		\caption{\label{fig:2q_nonlinearities}Illustration of non-monotonicity of SBS distance from the self-evolution of the environment parameter $\alpha_2$ and environmental mixedness (noise) $p$.}
	\end{figure}

	We have performed the calculation of the SBS distance~\eqref{eq:targetOptBasis} for the case with $\alpha_1=\alpha_3=0$ as a function of self-evolution of the environment parameter $\alpha_2$ and environmental mixedness, or \textit{noise}, parameter $p$ for various C-INOT imperfection parameter $\theta$. The results are shown at Fig.~\ref{fig:2qs}. For better readibility we show their marginal values for $p=0$ at Fig.~\ref{fig:2q_alpha2s}, and for $\alpha_2 = 0$ at Fig.~\ref{fig:2q_ps}.
	
	We observe that for $\theta > 0$ there exist values of $\alpha_2$ that allow improving the SBS structure of the evolved state, thus the self-evolution can to some extent counter-act the interaction gate imperfections.
	
	For the perfect C-NOT depicted in Fig.~\ref{fig:2qa} we observe that for small $p$ the self-evolution has a destructive influence on the SBS formation. On the other hand, for large values of $p$ adding some self-evolution may improve the SBS structure. It reveals, that for $p$ close to $0.5$ the Hadamard basis is the actual optimal basis for SBS formation. For $\alpha_2 \approx 1.5$ we observe a surprising phenomenon, that increasing the environmental mixedness may also improve the SBS formation.
	
	A similar situation of non-monotonicity in both $\alpha_2$ and $p$ can be clearly noticed in Figs~\ref{fig:2qb} and~\ref{fig:2qc} refering to imperfect C-NOT with $\theta = \pi /8$ and $\theta = \pi  /4$, respectively. For small $\alpha_2 \approx 0$ with increasing environmental mixedness the optimal SBS basis approaches the actual computational basis.
	
	For large imperfections of C-NOT, with $\theta = 0.9 \pi / 2$, Fig.~\ref{fig:2qd}, we see, that the SBS is being destroyed by noise in a monotonic way, but non-monotonicity in $\alpha_2$ shows that the state is closest to SBS for $\alpha_2 \approx 1.5$.
	
	To better illustrate the non-monotonic phenomena we depicted the marginal cases in Fig.~\ref{fig:2q_alpha2s}, where we show the SBS distance depending on $\alpha_2$, and in Fig.~\ref{fig:2q_ps}, where the dependence on $p$ is plotted.
	
	
\begin{figure}
		\centering   
		\subfigure[$\theta = 0$]{\label{fig:2qaa3}\includegraphics[width=60mm]{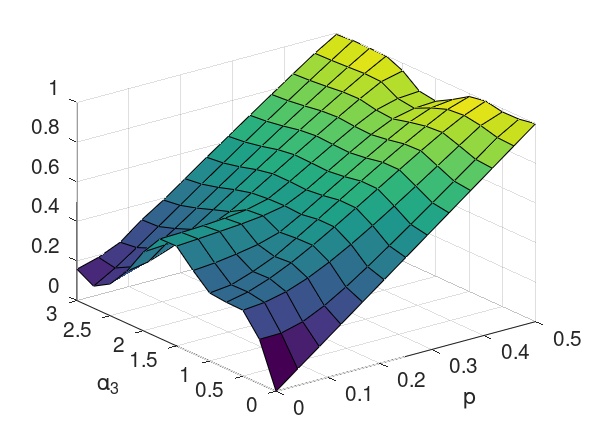}}
		\subfigure[$\theta = \pi / 8$]{\label{fig:2qba3}\includegraphics[width=60mm]{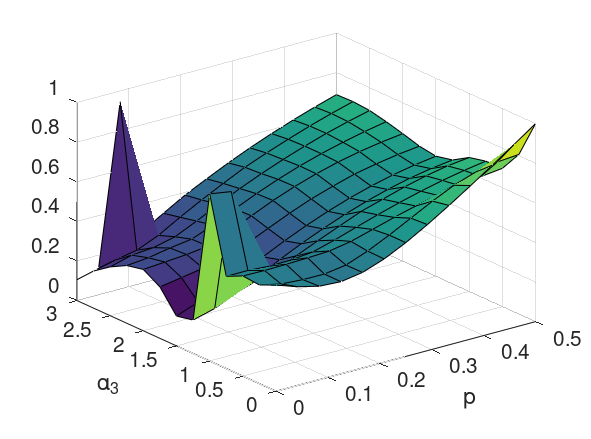}}
		\caption{\label{fig:2qsa3}SBS distance for intraction with C-INOT for various gate imperfection parameter $\theta$ with $2$ environmental qubits. Visible the dependence of the optimal environment mixedness $p$ on the value of the inter-environmental-evolution stength $\alpha_3$ for the Hamiltonian $H_3 = 2 \openone_2 \otimes \sigma_Z \otimes \sigma_Z$ instead of~\eqref{eq:H_3}.}
	\end{figure}

	For the sake of completeness, let us consider another form of the interaction between the two environmental qubits, that is the neighbour-neighbour interaction $2 \openone_2 \otimes \sigma_Z \otimes \sigma_Z$. We plot this dependence in Fig.~\ref{fig:2qsa3}. It can be seen, that the same non-monotonic pattern can be seen, as in Fig.~\ref{fig:2qs}.
	
	\subsection{Importance of the basis choice}
	
    In the present section, we shall illustrate the emergence of different indicator bases than the pointer basis in another way, rougher than the one performed in section 3.2. Namely, rather than performing full optimization, we perform a partial one, fixing the first one of the two anticipated bases (standard or Hadamard) and analyzing a specific parameter that will tell us which of the bases is closer to the optimum.
	
	To be more specific, in the optimization of the quantity ~\eqref{eq:targetOptBasis} we allowed for {\it any} SBS basis $\ket{\phi}$ of the observed system in the calculation of the minimal distance of $\rho_{S \mathcal{E}_1 \text{comp}}$ from the SBS set.
	
	Now, let us assume that the basis $\ket{\phi}$ is fixed, and the optimization is performed only over pure qubits $\ket{\chi}$ and $\tilde{p} \in [0,1]$. To this end, let us define the following subset of SBS states:
	\begin{equation}
	    \mathbb{S}_{\ket{\psi}} \equiv \left\{ \sigma : \sigma = \tilde{p} \proj{\psi} \otimes \proj{\chi} + (1-\tilde{p}) \proj{\psi^\perp} \otimes \proj{\chi^\perp}, \tilde{p} \in [0,1] \right\}.
	\end{equation}
	
	We define the distance $\mathcal{D}$ of the state $\rho$ from the set $\mathbb{S}$:
	\begin{equation}
	    \label{eq:basisDiff}
	    \mathcal{D} \left[ \rho, \mathbb{S} \right] \equiv \min_{\sigma \in \mathbb{S}} \TN{\rho - \sigma}.
	\end{equation}
	
\begin{figure}
		\centering   
		\subfigure[$\theta = 0$]{\label{fig:2qad}\includegraphics[width=60mm]{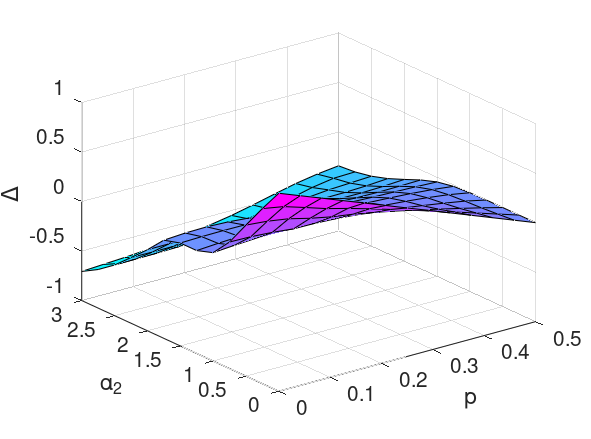}}
		\subfigure[$\theta = \pi / 8$]{\label{fig:2qbd}\includegraphics[width=60mm]{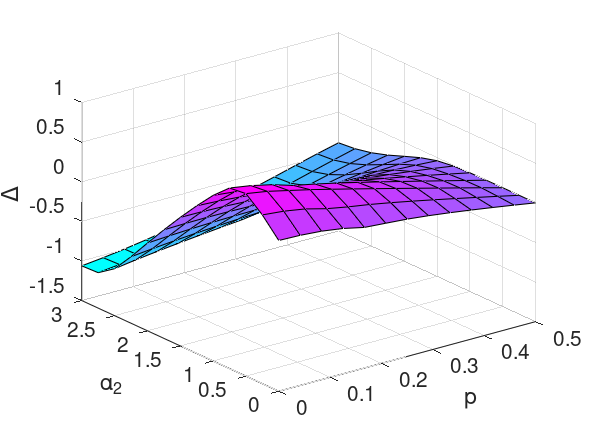}}
		
		\subfigure[$\theta = \pi / 4$]{\label{fig:2qcd}\includegraphics[width=60mm]{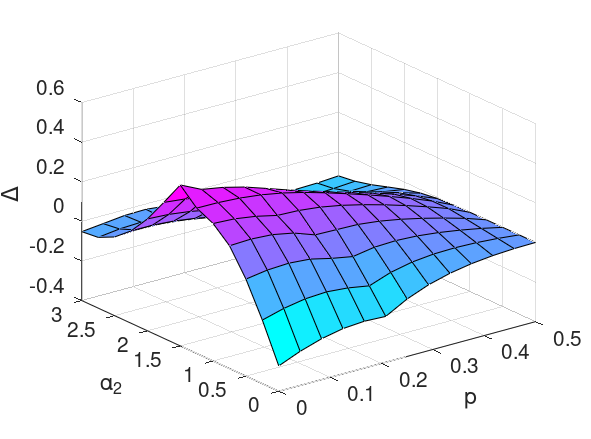}}
		\subfigure[$\theta = 0.9 \pi / 2$]{\label{fig:2qdd}\includegraphics[width=60mm]{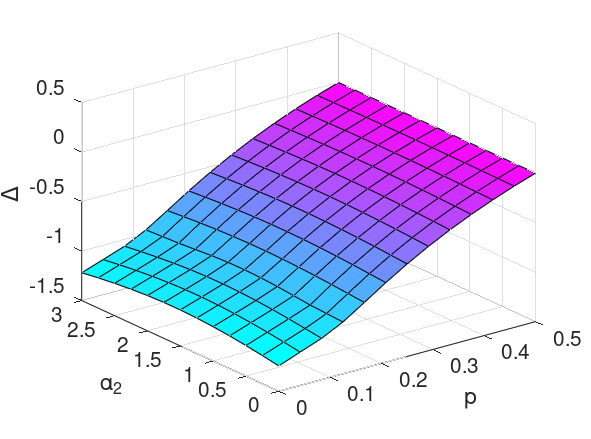}}
		\caption{\label{fig:2qs_diff} The difference $\Delta$, see~\eqref{eq:Delta}, of SBS distance for interaction with C-INOT various gate imperfection parameter $\theta$ with $2$ environmental qubits if the SBS is restricted to be in the Hadamard basis subtracted with the SBS distance if the SBS is restricted to be in the computational basis. The warmer color indicates that the evolved state $\rho_{S \mathcal{E}_1 \text{comp}}$ is closer to SBS in the computational basis, and the cooler color is in those regions, where the evolved state is closer to SBS in the Hadamard basis.}
	\end{figure}

	Now, we illustrate the difference between choices of different bases by comparing SBS distance if the basis of the SBS state is fixed to be either in the computational or in the Hadamard basis in \eqref{eq:optState}. To this end in Fig.~\ref{fig:2qs_diff} we plot the difference between the minimized SBS-distance in the latter basis subtracted the minimized SBS-distance in the former basis, \textit{viz.}
	\begin{equation}
	    \label{eq:Delta}
	    \Delta \equiv \mathcal{D} \left[ \rho_{S \mathcal{E}_1 \text{comp}}, \mathbb{S}_{\ket{+}} \right] - \mathcal{D} \left[ \rho_{S \mathcal{E}_1 \text{comp}}, \mathbb{S}_{\ket{0}} \right].
	\end{equation}

	It can be easily seen that even if we are not considering the optimal basis from~ \eqref{eq:optState-1}, but restrict to two simplest choices, the computational and Hadamard, the formation of the SBS structure favors either the former or the latter basis depending on the evolution and environment parameters, even though the pointer basis in quantum Darwinism sense doesn't change. One must remember that this is a very rough picture if compared to that of section 3.2. However, it shows that some tendencies concerning the information about the system encoded in the environment may still be identified despite the use of less computational effort.

	\section{Central Interaction: Optimization of Spectrum Broadcast Structure for $8$ environmental qubits}
	
	Next, we considered a case with a larger number $N_\text{env}$ of environmental qubits. In this case, we consider the broadcast Hamiltonian to be a sum
	\begin{equation}
		H_\text{int} = \sum_{i=1}^{N_\text{env}} H_\text{C-INOT}^i,
	\end{equation}
	where $H_\text{C-INOT}^i$ is defined by~\eqref{eq:Hc-maybe} with transformation over $i$-th environmental qubit controled by the central system. We consider only the self-evolution of separate environmental qubits, so this is a direct generalization of the three-qubits case with $\alpha_1 = \alpha_3 = 0$ and arbitrary $\alpha_2$. The self-evolution Hamiltonian is (in analogy to the 
	3-qubit case from Section 2.1 ):
	\begin{equation}
		\label{eq:H2moreQubits}
		H_{2}=\alpha_2 \sum_{i=1}^{N_\text{env}} \sigma_Z^i,
	\end{equation}
	where $\sigma_Z^i$ acts on $i$-th environmental qubit.
	
	We performed numerical calculations for an $8$-qubit environment. We assumed that $7$ of these qubits constitute the observer, with the last qubit being trace-out. In all cases in this section, we considered the optimal SBS basis.
	
\begin{figure}
		\centering   
		\subfigure[$\theta = 0$]{\label{fig:8a}\includegraphics[width=60mm]{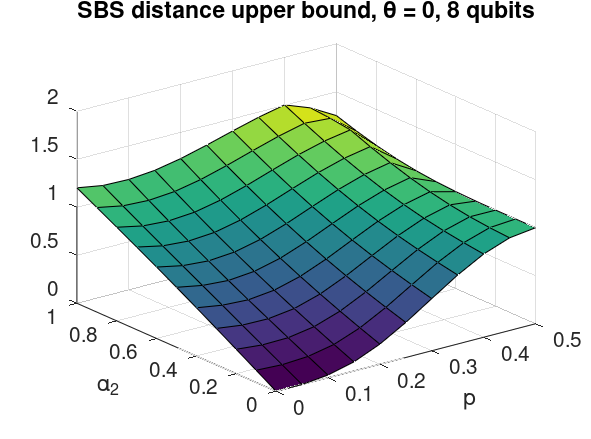}}
		\subfigure[$\theta = \pi / 8$]{\label{fig:8b}\includegraphics[width=60mm]{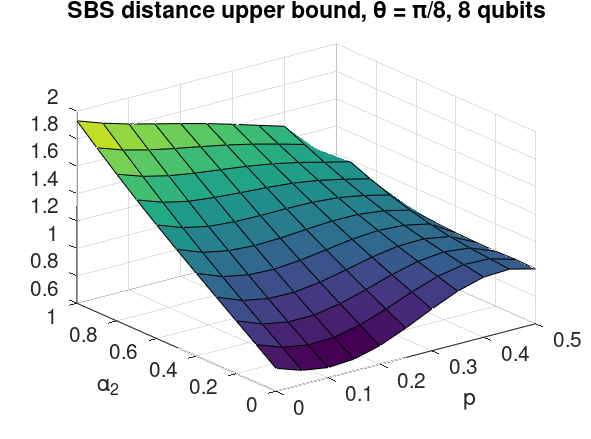}}
		
		\subfigure[$\theta = \pi / 4$]{\label{fig:8c}\includegraphics[width=60mm]{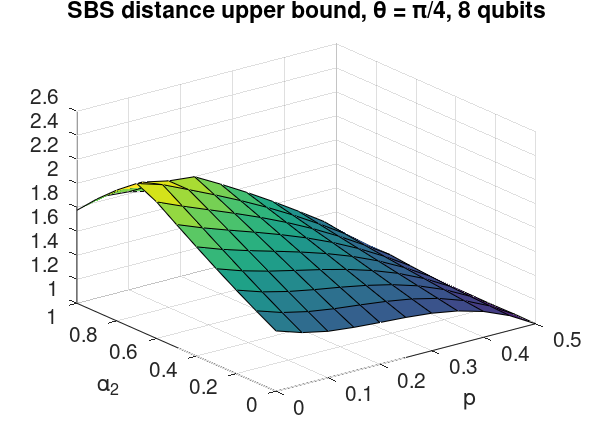}}
		\subfigure[$\theta = 0.9 \pi / 2$]{\label{fig:8d}\includegraphics[width=60mm]{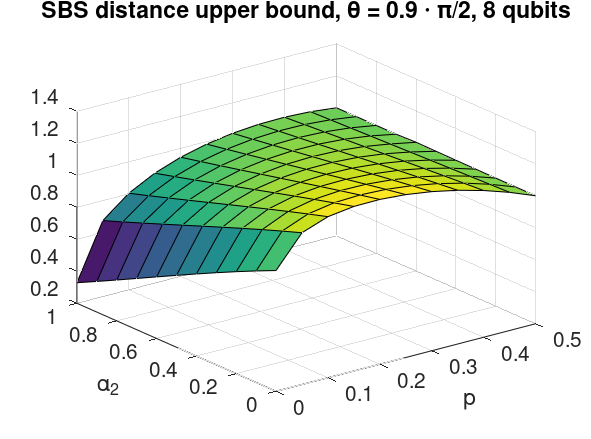}}
		\caption{\label{fig:8qs}SBS distance for C-INOT with various gate imperfection parameters $\theta$ with $8$ environmental qubits. Each value of $\theta$ refers to a different interaction between the central system and each of the environmental qubits, as given in~\eqref{eq:Uc-maybe}. The axis $\alpha_2$ describes the strength of the self-evolution of the environmental qubits, see~\eqref{eq:H2moreQubits}, and $p$ refers to the initial mixedness of the environmental qubits, see~\eqref{eq:initMixp}. The figure illustrate non-monotonic dependence of the \textit{upper bound}~\eqref{eq:OurPRLUpperBound} on the distance of the actually evolved state from the closes SBS state of the form~\eqref{eq:optState} on the parameters $\alpha_2$ and $p$. In particular, it can be seen that in many cases it is not the smallest value of mixedness, that leads to states closing (in an upper bound sense) to the SBS form, but the ``optimal'' environment mixedness $p$ depends on the value of the self-evolution strength $\alpha_2$.}
	\end{figure}

	The optimization of~\eqref{eq:targetOptBasis} for environments of dimension larger is much more difficult, thus we were not able to find the state~\eqref{eq:optStateHigher} exactly. Instead, we calculated the upper bound of~\cite{MonitMironowicz}, cf.~\eqref{eq:OurPRLUpperBound} to check, if the non-monotonic phenomena, that we observed for two qubits, can be expected to occur also in this case. The results of the numerical optimization are shown in Fig.~\ref{fig:8qs}. The calculated upper bounds suggest that there exists some regime of gate imperfection $\theta$, where both self-evolution and noise of the environment can improve the SBS structure like it was in the case of a two-qubit environment (see sections 3 and 4).
	
	We stress that the quantity~\eqref{eq:OurPRLUpperBound} of~\cite{MonitMironowicz} is only an upper bound, even though it can be applied to a state, and is easily computable. Note, that calculation of~\eqref{eq:OurPRLUpperBound} is purely algebraic, and doesn't require any optimization procedure. On the other hand, since it provides only an upper bound, the exact results may diverge from those obtained with the bound, yet the similarity of behavior of the plots obtained with the bound (see Fig.~\ref{fig:8qs}) is similar to those derived using optimization of an exact formula (see Fig.~\ref{fig:2qs}). It shows, that the upper bound is able to properly grasp the non-monotonic tendencies occurring in both scenarios.
	
	For perfect C-NOT, see eq.~\eqref{eq:Uc-maybe}, with $\theta = 0$, for majority values of $\alpha_2$ the SBS distance is gradually growing with increasing $p$, approaching value close to $1$ for the maximal mixedness $p \approx 0.5$. For $\alpha_2 \in [0,1]$ the SBS distance is also increasing for $p \approx 0$. Yet, for large values of $\alpha_2$ and $p$, a slightly non-monotonic behaviour is seen in $p$.
	
	For $\theta = \pi / 8$ a clear improvements in SBS formation with increasing $p$ can be seen at Fig.~\ref{fig:8b}, where also the optimal value of $\alpha_2$ is increasing with $p$. For $\alpha_2 \approx 0$ it can be observed, that the SBS is best formed for $p \approx 0.1$, that is also a surprising effect, confirming the previous observation that with self-evolution of environment, it is possible that more noised (mixed) initial environment is more suitable for SBS formation that the pure environment. Even stronger effect is visible at Fig.~\ref{fig:8c}. Still, it should be noted that in those cases the SBS distance upper-bound is very large, close to $1$, or even higher, so its behavior may serve only as a very preliminary suggestion regarding the behavior of the actual distance to SBS of the formed states, and as such, should be followed by tight analytical approximations in future.
	
	In the case of large imperfection of C-INOT, \textit{viz.} $\theta = 0.9 \pi / 2$, see Fig.~\ref{fig:8d}, a clear effect of improvement in SBS formation for increasing self-evolution of environment parameter $\alpha_2$ occurs, that is especially strong for small $p \approx 0$.

	\section{Non-central interaction for 8 qubits}
	
\begin{figure}
		\centering
		\includegraphics[width=0.7\linewidth]{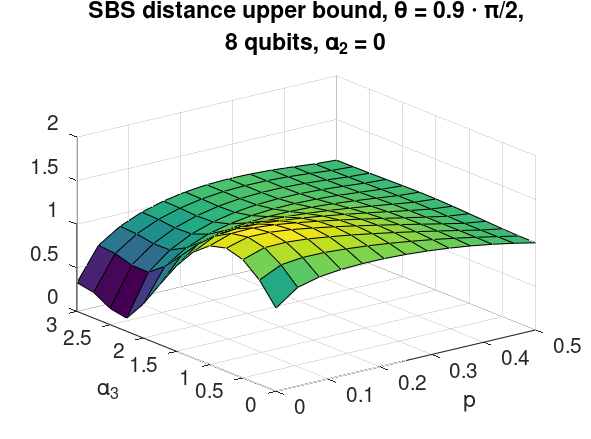}
		\caption{Upper bound on the distance to an SBS state for $8$-qubit environment and $\alpha_1 = \alpha_2 = 0$ as a function of neighbour-neighbour interaction~\eqref{eq:H_3many} stength $\alpha_3$ and mixedness $p$ of the environment.}
		\label{fig:upper8qubitsalpha3}
	\end{figure}

	Now, let us consider the case with interaction between environmental qubits of the following neighbour-neighbour form:
	\begin{equation}
		\label{eq:H_3many}
		H_3 = \alpha_3 \sum_{i=1}^{N_\text{env}} \sigma_Z^{i} \otimes \sigma_Z^{(i \text{ mod } N_\text{env} + 1)},
	\end{equation}
	where $N_\text{env}$ is the number of qubits in the environment, and $\sigma_Z^{i}$ acts on $i$-th qubit of the environment. We have calculated the upper bound~\cite{MonitMironowicz} for the case with $\alpha_1 = \alpha_2 = 0$ and $8$ qubits as a function of the imperfection of C-INOT parameter $\theta$ and environmental noise $p$. The results are show at Fig.~\ref{fig:upper8qubitsalpha3}. A strong non-monotonicity in $\alpha_3$ can be observed for low values of $p$, e.g in the case of $\theta = 0.9 \pi / 2$, where taking $\alpha_3 \approx 2$ can repair the effect of C-INOT imperfection.
	The analogous situation takes place for fixed $\alpha \approx 0.75 $ where the bound is decreasing with increasing initial noise for the region of $p \approx 0,1$. One should remember, however, that in this case the numerical values of the bound are high and the search for possible non-monotonous behaviour of the exact distance as a function of $p$ should be continued.
	
	\section{Conclusions and discussion}

We examined the aspects of the emergence of objective information in the dynamic physical scenario in low-dimensional qubit systems. More precisely we considered the non-perfect propagation of information from the system to the noisy environment with self-evolution, where the imperfect C-NOT gate \cite{Zurek_eavesdropping} is accompanied by the presence of the self-dynamics of the environment which – in general – may be in an initially mixed (thermal) state. We considered two different environments, the first composed of one observed and one unobserved qubit and 
the second one where there are seven observed qubits and one unobserved. In particular, we examined the analytical model three interacting qubits and we derived its objectivity parameters. 

    We have considered examples with the system in the Hadamard state and showed that if the imperfection of the C-NOT gate is known, the emergence of the objectivity – albeit with respect to a different basis than the one associated with the gate itself - can be enhanced by a carefully chosen environment self-dynamics which may be interpreted just as an external magnetic field.
    The numerical optimization shows that the quality of the spectrum broadcast structure formed during the interaction may be non-monotonic both in the speed of self-dynamics of the environment as well as its mixedness.  We interpret this phenomenon as the emergence of a new type of objectivity, which may be called a relaxed objectivization, since the statistics do not disclose any parameters of the initial state of the system, but present to the observer some new ones, generated during the complex dynamical process.

    We also discussed the case of 8 qubits of environment and numerical calculations support the general conjecture, that dynamics of the environment may help the emergence of objectivity to happen. It suggests that even if the imperfectness of the C-NOT is large enough to destroy objectivity in the standard scenario~\cite{Zurek_eavesdropping}, one may observe its ``comeback", as a kind of phase transition due to carefully tuned self-dynamics of the environment. 
    
    We believe, that the above concept of the relaxed objectivization is interesting in itself because it concerns the general question of whether the system is in fully classical relation with the environment in the philosophical, purely existential sense, namely that one is allowed to make a sensible claim that some of its property exists. In this sense, the present approach brings out the ontological aspect of emergent objectivity in a quantum world. 
    
    Possible cognitive and practical consequences of that go in two directions.
    First, if we are in the engineering paradigm we know that objectivity (technically represented here by SBS structure) makes system-environment composition useless for coherent quantum information processing. This may be important in experiments monitoring a general interaction of a given system with some mesoscopic environment including quantum memory and other coherent effects. In such cases, one should know methods to keep its state far from such an objective form. 
    Note, that the present analysis suggests that it can be done in a simple way, namely just by tuning an external magnetic field.
    Second, the present analysis may inspire several open questions concerning the possibility of the emergence of objectivity close to the original quantum Darwinism paradigm, yet more relaxed,  in some physical scenarios. 

    For instance, here we considered only the situation when the information was objectively "mirrored" in one environment (cf. \cite{Winter_Thermality}).
    Is it possible to observe the present, relaxed objectivity effect stable in time for a large number of environments like it was in the case of quantum Darwinism objectivity? If so, is it possible to find situations when, despite the "unfriendly circumstances" -  environment dynamics, noise, and deviation from the C-NOT gate interaction - the information about some parameters of the initial state of the system can still be easily retrieved from the environment? Another question would be, whether and when the present objectivised basis can be exploited to read out some well-defined parameters of whole dynamics. The original pointer basis was defined by a local element, ie. a system-environment interaction Hamiltonian. Concerning our case: does one need to know all the global dynamics, or are there cases when knowledge of some parameters of the global dynamics (and, maybe, its particular symmetries) is enough to determine our analog of the pointer basis?

    Note that for more than 2 environments the SBS structure is stronger than Strong Quantum Darwinism \cite{SQD}. However, the concept of generalized pointer basis in those dynamical scenarios where interaction Hamiltonian alone does not determine objectivity may be, in full analogy, defined for Strong Quantum Darwinism, since the latter is also agnostic to the physical mechanism leading to it. The corresponding system environment state satisfying SQD is of a quite general form
    $\varrho_{\mathcal{S} \mathcal{E'}} = \sum_i p_i \proj{\psi_i} \otimes \varrho_i^{\mathcal{E'}_1 \cdots \mathcal{E'}_N}$
    but with the special property. Namely, there must exist some isometries that act {\it locally} on the parts of environments 
    $ U_i^{\mathcal{E'}_1} \otimes \cdots \otimes U_i^{\mathcal{E'}_N}: {\mathcal{E'}_1}\otimes \cdots \otimes {\mathcal{E'}_N} \rightarrow {\mathcal{E}_1}{\mathcal{E"}_1}\otimes \cdots \otimes {\mathcal{E}_N\mathcal{E"}_N}$ and transform the state $\varrho_{\mathcal{S} \mathcal{E'}}$ into another state $\varrho_{\mathcal{S} \mathcal{EE"}}$ in such a way, that after tracing out the $\mathcal{E"}$ parts of the environment one gets the SBS state defined in \eqref{SBS-canonical} (the domains of the isometries involve also those degrees of freedom that carry possible correlations between different parts of environments but are irrelevant for objectivity).  If there are interactions between different parts of the environment it is likely that objectivity will be encoded in the above general SQD form due to correlations produced by the interactions. Searching for a generalized pointer basis in a dynamical system may be even more demanding, especially if the environment corresponds already to so-called macrofractions (see \cite{Qorigins}). In those cases most probably new analytical methods will be needed due to the complexity and numerical intractability of the problem. 

    Finally, the observed non-monotonicity of objectivity under the parameters of the two potentially "unfriendly" elements of the scenario - speed of environment dynamics and mixedness of its states seems counterintuitive. We believe that it needs further investigation in more complex models  -  both from the SBS as well as SQD perspective -  and may lead to some applications that are difficult to identify at the present, early stage of the analysis.

	\acknowledgments{The work is supported by the Foundation for Polish Science (IRAP project, ICTQT, contract no. 2018/MAB/5, co-financed by EU within Smart Growth Operational Programme). The numerical calculations we conducted using OCTAVE~6.1~\cite{OCTAVE}, and packages QETLAB~0.9~\cite{QETLAB} and Quantinf~0.5.1~\cite{quantinf}.}
	
	\abbreviations{Abbreviations}{
        The following abbreviations are used in this manuscript:\\
        
        \noindent
        \begin{tabular}{@{}ll}
            SBS & spectrum broadcast structure\\
            QD & quantum Darwinism\\
            SQD & strong quantum Darwinism\\
            C-NOT & controlled-NOT gate\\
            C-INOT & controlled imperfect-NOT gate\\
            $\mathcal{F}$ & fidelity\\
            $I(\mathcal{S}:\mathcal{E})$ & mutual information between the system and part of the environment\\
            $H(\cdot)$ & von Neumann entropy\\
            $\chi(\mathcal{S}:\mathcal{E})$ & Holevo information between $\mathcal{S}$ and $\mathcal{E}$
        \end{tabular}
    }
	
	\appendixtitles{no}
	\appendixstart
	\appendix
	
	\section[\appendixname~\thesection]{}
	\label{sec:AppA}
	
	Direct calculations show that $H_\text{TOTAL} = (\pi - \alpha_1) \openone_8 + M$, where $\openone_8$ is the 3-qubit identity operator, and $M \equiv \begin{bmatrix}	M_0 & 0 \\ 0 & M_1 \end{bmatrix}$ is a block diagonal matrix with $M_0$ and $M_1$ given by:
	\begin{subequations}
		\begin{equation}
			M_0 \equiv
			\begin{bmatrix}
				\xi_1 & 0 & 0 & 0 \\
				0 & \xi_2 & 0 & 0 \\
				0 & 0 & \xi_2 & 0 \\
				0 & 0 & 0 & \xi_3
			\end{bmatrix},
		\end{equation}
		\begin{equation}
			\label{eq:M_1}
			M_1 \equiv
			\begin{bmatrix}
				-y & -x/2 & -x/2 & 0 \\
				-x/2 & 0 & 0 & -x/2 \\
				-x/2 & 0 & 0 & -x/2 \\
				0 & -x/2 & -x/2 & y
			\end{bmatrix},
		\end{equation}
	\end{subequations}
	where we denote:
	\begin{subequations}
		\begin{equation}
			\xi_1 \equiv -\pi + 2\alpha_1 +2\alpha_2 +4\alpha_3,
		\end{equation}
		\begin{equation}
			\xi_2 \equiv -\pi +2\alpha_1 -2\alpha_3,
		\end{equation}
		\begin{equation}
			\xi_3 \equiv -\pi +2\alpha_1 -2\alpha_2,
		\end{equation}
		\begin{equation}
			x \equiv \pi \cos(\theta),
		\end{equation}
		\begin{equation}
			y \equiv \pi \sin(\theta) - 2\alpha_2 + 2\alpha_3.
		\end{equation}
	\end{subequations}
	We will often use the following term:
	\begin{equation}
		w \equiv \sqrt{x^2+y^2}.
	\end{equation}
	
	Calculating eigendecomposition of~\eqref{eq:M_1} we get that $M_1 = U \cdot D \cdot U^\dagger$, where $D$ is the diagonal matrix with elements $(0,0,w,-w)$, and unitary $U$ is given by:
	\begin{equation}
		U \equiv (1/2)
		\begin{bmatrix}
			0 & \sqrt{2} x/w & (w-y)/w & (w+y)/w \\
			\sqrt{2} & -\sqrt{2} y/w & -x/w & x/w \\
			-\sqrt{2} & -\sqrt{2} y/w & -x/w & x/w \\
			0 & -\sqrt{2} x/w & (w+y)/w & (w-y)/w
		\end{bmatrix}.
	\end{equation}
	Using this formula we can calculate $V \equiv \exp(-i t M)$ to be block diagonal with blocks $V_0$ and $V_1$, where $V_0$ is the diagonal matrix with elements $(u_1, u_2, u_2, u_3)$,
	\begin{equation}
	    u_i \equiv \exp(-i t \xi_i),
	\end{equation}
	and $V_1 = R + i Q$, with $R$ and $Q$ defined as follows:
	\begin{subequations}
		\begin{equation}
			R \equiv
			\begin{bmatrix}
				r_1 & r_2 & r_2 & r_3 \\
				r_2 & r_4 & r_3 & -r_2 \\
				r_2 & r_3 & r_4 & -r_2 \\
				r_3 & -r_2 & -r_2 & r_1
			\end{bmatrix},
		\end{equation}
		\begin{equation}
			Q \equiv
			\begin{bmatrix}
				-q_1 & q_2 & q_2 & 0 \\
				q_2 & 0 & 0 & q_2 \\
				q_2 & 0 & 0 & q_2 \\
				0 & q_2 & q_2 & q_1
			\end{bmatrix},
		\end{equation}
	\end{subequations}
	where
	\begin{equation}
		\begin{aligned}
			& r_1 \equiv 0.5 (x^2 + (w^2+y^2) \cos(t w))/w^2, \\
			& r_2 \equiv -0.5 x y (1-\cos(t w))/w^2, \\
			& r_3 \equiv -0.5 x^2 (1-\cos(t w))/w^2, \\
			& r_4 \equiv 0.5 (x^2 \cos(t w) + w^2+y^2)/w^2,
		\end{aligned}
	\end{equation}
	and
	\begin{equation}
		\begin{aligned}
			& q_1 \equiv -y \sin(t w)/w, \\
			& q_2 \equiv 0.5 x \sin(t w)/w.
		\end{aligned}
	\end{equation}
	One can check by direct calculations that the following identities hold:
	\begin{equation}
		\label{eq:rqIds}
		\begin{aligned}
			& r_1^2 + q_1^2 - r_4^2 = 0, \\
			& q_2^2 + r_2^2 + r_3^2 + r_3 = 0, \\
			& r_1 r_2 + r_2 r_3 - q_1 q_2 + r_2 = 0, \\
			& q_1 r_2 + q_2 r_1 - q_2 r_3 - q2 = 0, \\
			& r_4 - r_3 - 1 = 0, \\
		\end{aligned}
	\end{equation}
	and that $r_3 \in [-1,0]$ and $r_4 \in [0,1]$.

	\section[\appendixname~\thesection]{}
	\label{sec:AppB}
	
	Using the notation of Appendix~\ref{sec:AppA}, direct calculations show that for $\Gamma$ defined in~\eqref{eq:Gamma}~we have:
	\begin{equation}
		\Gamma = p \sqrt{s_1 + 2 \Re{(u_1 u_2^{*} s_2)} } + (1-p) \sqrt{s_1 + 2 \Re{(u_2 u_3^{*} s_2)} },
	\end{equation}
	where $\Re$ is the real part of a number, and
	\begin{subequations}
		\begin{equation}
			s_1 \equiv (p^2 + (1-p)^2) \cdot r_4,
		\end{equation}
		\begin{equation}
			s_2 \equiv p (1-p) \cdot \left( (r_1 + \iu q_1) r_4  - (r_2- \iu q_2)^2 \right).
		\end{equation}
	\end{subequations}

	The states $\varrho_0$ and $\varrho_1$, obtained by conditioning upon the system state in the computational basis and tracing out the second environmental qubit, are equal
	\begin{subequations}
		\label{eq:obss}
		\begin{equation}
			\varrho_0 =
			\begin{bmatrix}
				p & 0 \\
				0 & 1-p
			\end{bmatrix},
		\end{equation}
		\begin{equation}
			\varrho_1 =
			\begin{bmatrix}
				1 + p (2 r_4-1) - r_4 & (1-2 p) (r_2 + i q_2) \\
				(1-2 p) (r_2 - i q_2) & p (1 - 2 r_4) + r_4
			\end{bmatrix}.
		\end{equation}
	\end{subequations}

	From the above it follows that for
	\begin{equation}
		\mu = -p (1-p) (2 r_4 - 1)+0.5 r_4,
	\end{equation}
	we have
	\begin{equation}
		\label{eq:insideB}
		\sqrt{\varrho_0} \varrho_1 \sqrt{\varrho_0} - \mu \openone_2 = 
		\begin{bmatrix}
			(p-0.5) r_4 & \sqrt{p (1-p)} (1-2 p) (r_2+i q_2) \\
			\sqrt{p (1-p)} (1-2 p) (r_2-i q_2) & -(p-0.5) r_4
		\end{bmatrix}.
	\end{equation}
	
	Using~\eqref{eq:rqIds} we get that the eigenvalues of~\eqref{eq:insideB} are $\pm \nu$, where
	\begin{equation}
		\nu = 0.5 \Ab{1-2 p} \sqrt{r_4 (r_4 - 4 (p-p^2) \cdot (r_4-1))}.
	\end{equation}
	Thus, we have the following closed form for~\eqref{eq:B}:
	\begin{equation}
		\label{eq:Bclosed}
		\mathcal{F}(\varrho_0, \varrho_1) = \sqrt{\mu+\nu} + \sqrt{\mu-\nu}.
	\end{equation}

	\section[\appendixname~\thesection]{} 
	\label{sec:AppC}
	
	The state of an environmental qubit is given as the average of~\eqref{eq:obss}, so it equals
	\begin{equation}
		\label{eq:qubit_state}
		\begin{bmatrix}
			0.5 - r_4 (0.5 - p) & 0.5 (1-2 p) (r_2 + i q_2) \\
			0.5 (1-2 p) (r_2 - i q_2) & 0.5 + r_4 (0.5 - p)
		\end{bmatrix}.
	\end{equation}
	Now, let us consider how close is this state to the Gibbs state, in particular directly after the short-term C-INOT interaction, \textit{i.e.} for $t = 1$? 
	
	Since the Hamiltonians~\eqref{eq:Hs} are diagonal in computational basis and proportional to $\sigma_Z = \begin{bmatrix} 1 & 0 \\ 0 & -1\end{bmatrix}$, also the Gibbs state will be diagonal, with the second diagonal value greater or equal the first (for $\sigma_Z$ $\ket{1}$ is the ground state).
	
	Recall that from the form of the thermal environment~\eqref{eq:initMixp}, we have $p \in [0,0.5]$. One can check that $r_4 \in [0,1]$, so $r_4 (0.5 - p) \geq 0$. Thus the second diagonal term of~\eqref{eq:qubit_state} is greater or equal the first, and so the trace distance of the state~\eqref{eq:qubit_state} from the closest thermal state is given by
	\begin{equation}
		\TN{\begin{bmatrix}
				0 & 0.5 (1-2 p) (r_2 + i q_2) \\
				0.5 (1-2 p) (r_2 - i q_2) & 0
		\end{bmatrix}},
	\end{equation}
	where $\TN{\cdot}$ denotes the trace norm. This is equal to
	\begin{equation}
		\label{eq:thermal_distance}
		\Ab{1-2 p} \sqrt{r_2^2 + q_2^2} = \Ab{(1-2 p)} \sqrt{-r_4^2+r_4}.
	\end{equation}
	Direct calculations using~\eqref{eq:rqIds} show that~\eqref{eq:thermal_distance} can be rewritten as
	\begin{equation}
	    \label{eq:thermal_distance_simply}
	    \Ab{0.5-p} x \sqrt{1 - \cos\left(t \sqrt{x^2+y^2}\right)} \sqrt{\left(1+\cos\left(t \sqrt{x^2+y^2})\right)\right) x^2 + 2 y^2} /(x^2+y^2).
	\end{equation}
	
	Note, that for fixed $p$ and $\theta$ the value of~\eqref{eq:thermal_distance} is a function of $r_4$ that depends on the difference $\alpha_2 - \alpha_3$. The same holds for $\mathcal{F}(\varrho_0, \varrho_1)$, cf.~\eqref{eq:Bclosed}, as $\mu$ and $\nu$ are also functions of $p$, $\theta$ and $r_4$, and so there is a direct interplay between those two phenomena, the orthogonalization of observables and thermalization. 

	\begin{adjustwidth}{-\extralength}{0cm}
		
		\reftitle{References}
		
		

\begin{thebibliography}{999}
			\bibitem{Lands} Landsman, N.P. Between classical and quantum.
			Part of Philosophy of physics. {\bf 2007}, 417-553.    quant-ph/0506082 . DOI:  10.1016/B978-044451560-5/50008-7
			
			\bibitem{NoncomBarnum} H. Barnum, H.; Caves, C. M.;  Fuchs, C. A.; Jozsa, R.; Schumacher, B.  Noncommuting mixed states cannot be broadcast. Phys.Rev.Lett. {\bf 1996}, 76, 2818-2821. DOI: 10.1103/PhysRevLett.76.2818
			
			\bibitem{noBrodcastPiani} Piani, M.; Horodecki, P.; Horodecki R. No-Local-Broadcasting Theorem for Multipartite Quantum Correlations. Phys. Rev. Lett. {\bf 2008}, 100, 090502. DOI:https://doi.org/10.1103/PhysRevLett.100.090502
			
			\bibitem{Zurek 2009} Zurek, W.H. Quantum Darwinism. Nat. Phys. {\bf 2009}, 5, 181. doi: 10.1038/nphys1202.
			
			\bibitem{Zeh70} Zeh, H. D.  On the interpretation of measurement in quantum theory, Found. Phys. {\bf1970}, 1, 69. DOI:10.1007/BF00708656.
			
			\bibitem{Zurek81} Zurek, W. H. Pointer basis of quantum apparatus: Into what mixture does the wave packet collapse?. Physical Review D {\bf 1981 }, 24(6), 1516.
			
			\bibitem{ZurekRMP} Zurek, W. H. Decoherence, einselection, and the quantum origins of the classical. Rev. Mod. Phys. 2003, 75, 715 , DOI:https://doi.org/10.1103/RevModPhys.75.715.
			
			\bibitem{Joos2003} Joos, E.; Zeh, H. D.; Kiefer, C. ;  Giulini, D.; Kupsch, J.; Stamatescu, I.-O.;  Decoherence and the Appearance of a Classical World in Quantum Theory, 2nd Edition, Springer, New York, 2003.
			
			\bibitem{SchlosshauerRMP} Schlosshauer, M.  Decoherence, the measurement problem, and interpretations of quantum mechanics. Rev. Mod. Phys. {\bf 2005}, 76, 1267. DOI: https://doi.org/10.1103/RevModPhys.76.1267.
			
			\bibitem{Grangier} Auffeves, A.; Philippe Grangier, P. Recovering the quantum formalism from physically realist axioms. Scientific Reports {\bf 2017}, 7, 43365. DOI:10.1038/srep43365.
			
			\bibitem{MonitMironowicz} Mironowicz, P.; Korbicz, J.; Horodecki P. Monitoring of the process of system information broadcasting in time. Phys. Rev. Lett. {\bf 2017}, 118, 150501 DOI: 10.1103/PhysRevLett.118.150501.
			
			\bibitem{EnvirWitness} Ollivier, H.;  Poulin, D.; Zurek, W. H. Objective properties from subjective quantum states: Environment as a witness. Phys. Rev. Lett. {\bf 2004}, 93, 220401. DOI: 10.1103/PhysRevLett.93.220401.

			\bibitem{Qorigins} Horodecki, R.; Korbicz, J. K.; Horodecki P. Quantum origins of objectivity. Phys. Rev. A {\bf 2015}, 91, 032122.  DOI: 10.1103/PhysRevA.91.032122.
			
			\bibitem{Zurek_eavesdropping} Touil, A.; Yan, B.; Girolami, D.; Deffner, S.; Zurek W. H. Eavesdropping on the Decohering Environment: Quantum Darwinism, Amplification, and the Origin of Objective Classical Reality Phys.Rev.Lett. {\bf 2022}, 128, 010401. DOI: 10.1103/PhysRevLett.128.010401 .
			
			\bibitem{QD_struct.envir} Pleasance G.; Garraway B. M. Application of quantum Darwinism to a structured environment. Phys. Rev. A {\bf 2017}, 96, 062105. DOI: https://doi.org/10.1103/PhysRevA.96.062105 
			
			\bibitem{Le_QD_SBS} Le, T.P.;  Olaya-Castro, A. Objectivity (or lack thereof): Comparison between predictions of quantum Darwinism and spectrum broadcast structure. Phys. Rev. A 2018, 98, 032103. DOI:https://doi.org/10.1103/PhysRevA.98.032103
			
			\bibitem{Korbicz_road} Korbicz, J.K.; Roads to objectivity: Quantum Darwinism, Spectrum Broadcast Structures, and Strong quantum Darwinism -- a review. Quantum 
			{\bf 2021 }, 5, 571. DOI: 10.22331/q-2021-11-08-571.
			
			\bibitem{Palma_WitnessObjComputer} Chisholm, D.A.; Guillermo García-Pérez, D.A.;  Rossi, M.A.C.; Maniscalco, S.;  Palma, G. M. Witnessing Objectivity on a Quantum Computer {\bf 2021}, arXiv:2110.06243. 
			
			\bibitem{Blurred}  Le, T.P.; Mironowicz, P.; Horodecki , P. Blurred quantum Darwinism across quantum reference frames. Phys. Rev. A {\bf 2020}, 102, 062420. DOI:10.1103/physreva.102.062420.
			
			\bibitem{Tuziemski} Tuziemski, J. Decoherence and information encoding in quantum reference frames  {\bf 2020 }, arXiv: 2006.07298v2.
			
			\bibitem{Winter_Thermality}   Le, T.P.;  Winter, A.; Adesso G. Thermality versus objectivity: can they peacefully coexist? Entropy {\bf 2021}, 23(11), 1506. DOI:10.3390/e23111506.
			
			\bibitem{SQD} Le, T.P.;  Olaya-Castro, A. Strong Quantum Darwinism and Strong Independence is equivalent to Spectrum Broadcast Structure. Phys. Rev. Lett. {\bf 2019}, 122, 010403. DOI:10.1103/PhysRevLett.122.010403.
			
			\bibitem{CommentSQD} Feller, A.;  Roussel, B.; Frérot, I.; Degiovanni P. Comment on ``Strong Quantum Darwinism and Strong Independence are Equivalent to Spectrum Broadcast Structure". Phys. Rev. Lett. {\bf 2021}, 126, 188901. DOI:10.1103/PhysRevLett.126.188901.
			
			\bibitem{ReplaySQD}   Le, T.P.; Olaya-Castro, A.  Reply to Comment on ``Strong Quantum Darwinism and Strong Independence are Equivalent to Spectrum Broadcast Structure". {\bf 2021}, arXiv:2101.10756.
			
			\bibitem{Witnessing-non-object}   Le, T.P.; Olaya-Castro, A. Witnessing non-objectivity in the framework of strong quantum Darwinism. Quantum Sci. Technol. {\bf 2020}, 5, 045012. DOI: 10.1088/2058-9565/abac4e.
			
			\bibitem{FuchsFidelity} Fuchs, C. A.; van de Graaf, J., Cryptographic distinguishability measures for quantum-mechanical states, IEEE Trans. on Inf. Theor. 45, 1216 (1999).
			
			\bibitem{gradientMethod} Polak, E. Optimization : Algorithms and Consistent Approximations. Springer-Verlag. ISBN 0-387-94971-2 (1997).
			
			
			
			\bibitem{Zurek09} Zwolak, M.; Quan, H. T.; Zurek, W. H. Quantum Darwinism in a mixed environment. Physical Review Letters, {\bf 2009}, 103 (11), 110402.
			
			\bibitem{Roszak21}  Roszak, K.;  Korbicz, J. K.; Glimpse of objectivity in bipartite systems for nonentangling pure dephasing evolutions, Phys. Rev. A {\bf 2020} 101, 052120.
			
			
			\bibitem{OCTAVE} Eaton, J. W.; Bateman, D., Hauberg, S.; Wehbring, R. {GNU Octave} version 6.1.0 manual: a high-level interactive language for numerical computations, \url{https://www.gnu.org/software/octave/doc/v6.1.0/}, 2020.
			
			\bibitem{QETLAB} Johnston, N. \textit{{QETLAB}: A {MATLAB} toolbox for quantum entanglement, version 0.9}, \url{http://qetlab.com}, 2016.
			
			\bibitem{quantinf} Cubitt, T. Quantinf Matlab Package, version 0.5.1, \url{https://www.dr-qubit.org/matlab.html}, 2013.
			
		\end{thebibliography}
		

	\end{adjustwidth}
\end{document}